\documentclass[showpacs, notitlepage, preprintnumbers, prc, twocolumn, floatfix]{revtex4-1}
\usepackage{amssymb}
\usepackage{amsmath}
\usepackage{graphicx}
\usepackage{bm}
\usepackage{hyperref}

\begin{document}

\title{How large is the Knudsen number reached in fluid dynamical simulations of ultrarelativistic heavy ion collisions? }
\author{H.\ Niemi${}^{a,b}$ and G.\ S.\ Denicol${}^{c}$}

\affiliation{$^{a}$Department of Physics, University of Jyv\"askyl\"a, P.O. Box 35 (YFL), 
FI-40014 University of Jyv\"askyl\"a, Finland} 
\affiliation{$^b$Helsinki Institute of Physics, P.O.Box 64, FI-00014 University of
Helsinki, Finland} 
\affiliation{$^{c}$Department of Physics, McGill University, 3600 University Street,
Montreal, Quebec, H3A 2T8, Canada}

\begin{abstract}
We investigate the applicability of fluid dynamics in ultrarelativistic heavy 
ion (AA) collisions and high multiplicity proton nucleus (pA) collisions. In 
order for fluid dynamics to be applicable the microscopic and macroscopic 
distance/time scales of the system have to be sufficiently separated. The degree 
of separation is quantified by the ratio between these scales, usually referred 
to as the Knudsen number. In this work, we calculate the Knudsen numbers reached 
in fluid dynamical simulations of AA and pA collisions at RHIC and LHC energies. 
For this purpose, we consider different choices of shear viscosity 
parametrizations, initial states and initialization times. We then estimate the 
values of shear viscosity for which the fluid dynamical description of 
ultrarelativistic AA and pA collisions breaks down. In particular, we study how 
such values depend on the centrality, in the case of AA collision, and 
multiplicity, in the case of pA collision. We found that the maximum viscosity 
in AA collisions is of the order $\eta/s \sim 0.1 \ldots 0.2$,  which is similar 
in magnitude to the viscosities currently employed in simulations of heavy ion 
collisions. For pA collisions, we found that such limit is significantly lower, 
being less than $\eta/s=0.08$ 
\end{abstract}

\maketitle


\section{Introduction}

The main goal of the experimental heavy ion collision program at RHIC and
LHC is to verify the existence of a nearly thermal quark-gluon plasma (QGP),
as well as to determine its properties. Currently, the several
collaborations participating in these experiments are able to provide a
wealth of experimental data on the transverse momentum of hadrons, leptons
and photons produced in the collisions. However, interpreting such findings
and extracting the properties of the QGP from them is a complicated task,
requiring a good theoretical understanding of the dynamics of the system
produced in the reaction.

Relativistic fluid dynamics is currently the main theoretical tool used to
describe the spacetime evolution of the bulk matter formed in
ultrarelativistic heavy ion collisions. Such approach has been particularly
successful in predicting and understanding the azimuthal asymmetries of the
transverse momentum spectra of charged hadrons~\cite{Heinz:2013th,
Huovinen:2013wma}. In this approach, such azimuthal asymmetries emerge from
a fluid-dynamical response to the azimuthally inhomogeneous initial geometry
of the matter produced. Currently, there are several studies indicating that
the effective shear viscosity to entropy density ratio $\eta /s$ required to
describe the data must be small, of the order of $\eta /s\sim 0.1\ldots 0.2$ 
\cite{Romatschke:2007mq,Luzum:2008cw,Schenke:2010rr,Gale:2012rq,Song:2010mg,
Song:2011qa,Shen:2010uy,Bozek:2009dw,Bozek:2012qs, Niemi:2011ix,
Niemi:2012ry, Paatelainen:2013eea}.

On the other hand, such collisions occur under extreme conditions that may
even question the validity of a fluid-dynamical description. The system is
created in a very small volume and lives for a very short time, rendering
the applicability of fluid dynamics questionable. Recently, the situation
was made more extreme by the inclusion of event by event fluctuations in the
initial state of the system~\cite{Andrade:2006yh, Alver:2010gr}, which serve
to increase the size of its energy density and flow velocity gradients. 
While in previous fluid-dynamical models such gradients were of the
order of the system size ($\sim 10$ fm), now the fluctuations are thought to
be sub-fermionic \cite{Dumitru:2012yr, Schenke:2012wb}. If fluid dynamics is 
in fact able to describe the time evolution of energy density fluctuations 
on such small distance scales, it may be even applicable to model the 
dynamics of proton-nucleus collisions \cite{Qin:2013bha, Werner:2013ipa, Bozek:2014era}.

For this reason, it is important to study the domain of validity of fluid
dynamics in ultrarelativistic heavy ion collisions. The applicability of
fluid dynamics is traditionally quantified in terms of the Knudsen number, a
ratio between the microscopic and macroscopic length scales involved in the
problem. A small Knudsen number implies a wide separation between the
microscopic and macroscopic distance/time scales making it possible for the
system to approach local thermodynamic equilibrium in macroscopic volumes. A
large Knudsen number usually results in a strong deviation from local
thermodynamic equilibrium, rendering the applicability of fluid dynamics
doubtful. So far, a detailed study of the values of Knudsen numbers that may
be reached in the realistic fluid-dynamical modeling of heavy ion collisions
has not been performed.

In this work, we estimate the magnitude of the (local) Knudsen numbers that
can be reached in the fluid-evolution of the quark-gluon plasma formed in
heavy ion (AA) collisions, and in proton-nucleus (pA) collisions. We
consider several choices of gradients to quantify the macroscopic
length/time scales: 1) the expansion rate, 2) energy density gradient, 3)
the shear tensor, 4) the vorticity tensor, and 5) the proper time derivative
of the flow velocity. We then investigate the applicability of fluid
dynamics in AA and pA collisions at RHIC and LHC energies. Here, we do not
consider fluctuations in the initial energy density profiles. Rather, we
study how the applicability of fluid dynamics depends on the scale of
density and velocity fluctuations by varying the system size. This is done
so by considering several centrality classes, for the case of AA collisions,
and several multiplicities for pA collisions. We then study the magnitude of
the viscosity that the fluid-dynamical modeling can handle before it breaks
down.

This paper is organized as follows. In Sec.~\ref{sec:fluid} we shortly
introduce the theory of transient fluid dynamics, and in Sec.~\ref{sec:knudsen} 
we discuss the conditions for which this theory is applicable.
The details of the fluid dynamical modeling of AA and pA collisions are
specified in Sec.~\ref{sec:urhic}. In Sec.~\ref{sec:results} we show the
main results of this paper. Finally, in Sec.~\ref{sec:conclusions} we
discuss the implications of our results and make our concluding remarks.

We use natural units $\hbar = c = k_{B} = 1$, and the sign convention of the
metric tensor is $g_{\mu\nu }=\mathrm{diag}\,(+,-,-,-)$.

\section{Relativistic Fluid dynamics}

\label{sec:fluid}

Fluid dynamics is an effective theory that describes the long distance, long
time limit of an underlying microscopic theory \cite{Landau}. In this case
the state of the fluid can be characterized solely by the densities and
currents associated to conserved quantities, such as the energy-momentum
tensor, $T^{\mu \nu }$, and net charge 4-currents, $N^{\mu }$ (e.g.\ the baryon
number 4-current in heavy ion collisions.) The dynamical evolution of $%
T^{\mu \nu }$ and $N^{\mu }$ is given by the continuity equations, 
\begin{equation}
\partial _{\mu }T^{\mu \nu }=0,\text{ }\partial _{\mu }N^{\mu }=0\text{,}
\label{eq:cons}
\end{equation}
which are completely general and follow solely from the fact that energy,
momentum, and charge are conserved on a microscopic level.

The energy-momentum tensor and charge 4-current can be tensor decomposed in
terms of the fluid 4-velocity, $u^{\mu }$, 
\begin{eqnarray}
T^{\mu \nu } &=&\varepsilon _{0}u^{\mu }u^{\nu }-\Delta ^{\mu \nu }\left(
P_{0}+\Pi \right) +\pi ^{\mu \nu }, \\
N^{\mu } &=&n_{0}u^{\mu }+q^{\mu },  \label{eq:Tmunu}
\end{eqnarray}%
where $\varepsilon _{0}$ is the energy density, $P_{0}$ the thermodynamic
pressure, $n_{0}$ the net-charge density, $\Pi $ the bulk viscous pressure, 
$q^{\mu }$ the net-charge diffusion 4-current, and $\pi ^{\mu \nu }$ the
shear-stress tensor. For the sake of convenience, we further introduced the
projection operator onto the 3-space orthogonal to the velocity, $\Delta
^{\mu \nu }=g^{\mu \nu }-u^{\mu }u^{\nu }$. It is also convenient to define
a double projection operator, that is symmetric and traceless, $\Delta
_{\alpha \beta }^{\mu \nu }=(\Delta _{\alpha }^{\mu }\Delta _{\beta }^{\nu
}+\Delta _{\beta }^{\mu }\Delta _{\alpha }^{\nu })/2-\Delta ^{\mu \nu
}\Delta _{\alpha \beta }/3$.

Since the conservation laws lead to only 5 equations of motion and conserved
currents, $T^{\mu \nu }$ and $N^{\mu }$, contain 14 degrees of freedom, the
system of equations is not yet closed and one still needs to provide 9
additional equations of motion. These would correspond to the time evolution
equations for the dissipative currents, $\Pi $, $q^{\mu }$, and $\pi ^{\mu
\nu }$.

In heavy ion collisions, one must model the evolution of the matter 
using a relativistic formulation of fluid-dynamics. Traditional
formulations, such as Navier-Stokes theory or the gradient expansion 
\cite{Chapman, Burnett}, do not work since their acausal nature \cite{his} leads
to intrinsic instabilities that render the theory useless for practical
purposes \cite{Denicol:2008ha,Pu:2009fj}. This problem is usually solved by
employing causal formulations of fluid dynamics, in which the dissipative
currents become independent dynamical variables that satisfy relaxation-type
equations \cite{Grad,IS}.

There are several formulations of causal fluid dynamics. In this work, we
employ the equations derived from kinetic theory in Refs.~\cite{IS,DKR,DNMR,
Denicol:2012es,Molnar:2013lta, Betz:2008me, Betz:2010cx, prd}. Such
equations include all the possible nonlinear terms that can appear in the
time evolution equations for the dissipative currents and in principle
increase the domain of applicability. For the purposes of this work, we
shall neglect the effects of bulk viscous pressure and net baryon number
diffusion, i.e., $\Pi =0=q^{\mu }$. Thus, all dissipative effects originate
only from the dynamics of the shear-stress tensor.

The equation of motion for the shear-stress tensor derived from kinetic
theory is \cite{DNMR} 
\begin{align}
\tau _{\pi }\dot{\pi}^{\left\langle \mu \nu \right\rangle }+\pi ^{\mu \nu }&
=2\eta \sigma ^{\mu \nu }+2\tau _{\pi }\pi _{\lambda }^{\left\langle \mu
\right. }\omega ^{\left. \nu \right\rangle \lambda }-\delta _{\pi \pi }\pi
^{\mu \nu }\theta  \notag \\
& -\tau _{\pi \pi }\pi ^{\lambda \left\langle \mu \right. }\sigma _{\lambda
}^{\left. \nu \right\rangle }+\varphi _{7}\pi ^{\lambda \left\langle \mu
\right. }\pi _{\lambda }^{\left. \nu \right\rangle }\;.  \label{eq:shear_eq}
\end{align}
Above we introduced the shear viscosity, $\eta $, shear relaxation time, $%
\tau _{\pi }$, and the remaining transport coefficients of the nonlinear
terms $\delta _{\pi \pi }$, $\tau _{\pi \pi }$, and $\varphi _{7}$. The
notation for the transport coefficients was taken from Ref.\ \cite{DNMR}. We
further introduced the shear tensor, $\sigma ^{\mu \nu }=$ $\nabla
^{\left\langle \mu \right. }u^{\left. \nu \right\rangle }$, the expansion
rate, $\theta =\nabla _{\mu }u^{\mu }$, and the vorticity tensor, $\omega
^{\mu \nu }=(\nabla ^{\mu }u^{\nu }-\nabla ^{\nu }u^{\mu })/2$, and used the
notation, $\dot{A}\equiv u^{\mu }\partial _{\mu }A$, $\nabla ^{\mu }\equiv
\Delta ^{\mu \nu }\partial _{\nu }$, and $A^{\left\langle \mu \nu
\right\rangle }=\Delta _{\alpha \beta }^{\mu \nu }A^{\alpha \beta }$.

In the 14-moment approximation and massless limit, it is possible to show
that \cite{DNMR,Molnar:2013lta} 
\begin{eqnarray}
\tau _{\pi } &=&5\frac{\eta }{\varepsilon _{0}+P_{0}}\text{ }, \\
\delta _{\pi \pi } &=&\frac{4}{3}\tau _{\pi }\text{ }, \\
\tau _{\pi \pi } &=&\frac{10}{7}\tau _{\pi }\text{ }, \\
\varphi_{7} &=&\frac{9}{70}P_{0}^{-1}
\text{ }.
\end{eqnarray}
In this work, we assume that the above expressions remain valid even for a
strongly interacting system, and use them in our simulations. In heavy ion
collisions, it is common to assume an effective shear viscosity that is
proportional to the entropy density, $\eta_{\mathrm{eff}}=a\times s$, where 
$s$ is the entropy density and $a$ is a proportionality coefficient. The
coefficient $a$ is usual extracted phenomenologically from heavy ion
collision experiments and may range from $a=0,\ldots ,0.2$. In this work, we
shall consider this effective viscosity, but also use temperature dependent
parametrizations of $\eta/s$.

\section{Knudsen number and validity of fluid dynamics}

\label{sec:knudsen}

As already mentioned, the validity of the fluid-dynamical description can
be quantified by the Knudsen number, \textrm{Kn}$=\ell _{\mathrm{micro}}/L_{\mathrm{macro}}$, 
where $\ell _{\mathrm{micro}}$ and $L_{\mathrm{macro}}$
are the microscopic and macroscopic distance/time scales, respectively. For
a dilute gas, $\ell _{\mathrm{micro}}$ is the mean free-path of the
particles and one can show that $\tau _{\pi }\sim \ell _{\mathrm{micro}}$.
For a strongly interacting system, the microscopic scale is not well known,
but we shall continue to assume that it is proportional to the shear
relaxation time, that is, 
\begin{equation}
\ell _{\mathrm{micro}}\sim \tau _{\pi }=5\frac{\eta }{\varepsilon _{0}+P_{0}}.
\end{equation}
We also note that the shear relaxation time is the only microscopic scale
that directly enters into Eq.~(\ref{eq:shear_eq}). The macroscopic scale
should be estimated from gradients of the macroscopic variables. Here, we
consider the following estimates for this scale 
\begin{eqnarray}
\frac{1}{L_{\mathrm{macro}}^{\theta }} &=&\theta \text{ ,}
\label{eq:ekspansion} \\
\frac{1}{L_{\mathrm{macro}}^{\varepsilon }} &=&\frac{1}{\varepsilon _{0}}
\sqrt{\nabla _{\mu }\varepsilon _{0}\nabla ^{\mu }\varepsilon _{0}}\text{ ,}
\label{eq:grade} \\
\frac{1}{L_{\mathrm{macro}}^{\sigma }} &=&\sqrt{\sigma _{\alpha \beta
}\sigma ^{\alpha \beta }}\text{ ,}  \label{eq:sigma} \\
\frac{1}{L_{\mathrm{macro}}^{\omega }} &=&\sqrt{\omega _{\alpha \beta
}\omega ^{\alpha \beta }}\text{ ,}  \label{eq:omega} \\
\frac{1}{L_{\mathrm{macro}}^{\dot{u}}} &=&\sqrt{\dot{u}_{\alpha }\dot{u}%
^{\alpha }}\text{ .}  \label{eq:udot}
\end{eqnarray}
That is, we only estimate the Knudsen number using different combinations of
gradients of velocity or the local energy density. The corresponding Knudsen
numbers are then given as $\mathrm{Kn}_{i}=\tau _{\pi }/L_{\mathrm{macro}}^{i}$. 
We note that in perfect fluid dynamics the scales \eqref{eq:grade}
and \eqref{eq:udot} are algebraically related, but the same does not hold in
viscous fluid dynamics where the difference depends also on the gradients of
the shear-stress tensor. As a matter of fact, in viscous fluid dynamics 
$L_{\mathrm{macro}}^{\dot{u}}$ and $L_{\mathrm{macro}}^{\varepsilon }$ give rise
to rather different macroscopic scales. For the sake of completeness we
listed all the possible macroscopic scales that could be constructed from the
gradients of velocity and of thermodynamic variables. However, in all the
fluid-dynamical simulations performed for the purposes of this paper, the
macroscopic scales $L_{\mathrm{macro}}^{\theta }$ and $L_{\mathrm{macro}}^{\varepsilon }$ 
were always considerably smaller than all the others.
Therefore, in order to obtain the domain of applicability of fluid dynamics,
it is enough to compare the microscopic scale only to $L_{\mathrm{macro}}^{\theta }$ 
and $L_{\mathrm{macro}}^{\varepsilon }$.

The range of Knudsen number values for which relativistic dissipative fluid
dynamics remains valid has been estimated by comparing fluid-dynamical
solutions to numerical solutions of the Boltzmann equation 
\cite{Molnar:2004yh, Huovinen:2008te, Molnar:2013lta, Bouras:2009nn,
Bouras:2010hm, Yano:2012mm, Denicol:2012vq, DKR}. Such studies were
performed considering dilute systems, for which the Boltzmann equation is
able to provide a reasonable description. Nevertheless, they should still be
able to provide some understanding about the range of applicability of fluid
dynamics in terms of the Knudsen number for other types of fluids. So far,
the best information on these limits come from comparing the solutions of
Israel-Stewart theory to numerical solutions of the Boltzmann equation in
simplified situations, such as (1+1)--dimensional shock problem~\cite{Bouras:2010hm} 
or (0+1)--dimensional Bjorken expansion~\cite{Huovinen:2008te}. In 
these situations it was found that a $\mathrm{Kn<0.5}$
is required to get a good agreement between Israel-Stewart theory and the
relativistic Boltzmann equation.

We note that the two tests mentioned above were done with Israel-Stewart
theory, which is not the exact fluid-dynamical limit of the Boltzmann
equation, see Refs.~\cite{DNMR, Denicol:2012vq}. Nevertheless, the fact that
the two very different tests give the same limit for $\mathrm{Kn}$, makes us
confident that $\mathrm{Kn=0.5}$ provides a reasonable estimate for the
limit of applicability of fluid dynamics. In this paper, we assume that this
limit can be applied to the Knudsen number defined from the smallest
macroscopic scale, that is, 
\begin{equation}
\mathrm{Kn}\equiv \frac{\tau _{\pi }}{L}\text{ , \ \ }L\equiv \min (L_{%
\mathrm{macro}}^{\theta },L_{\mathrm{macro}}^{\varepsilon })\text{ }.
\end{equation}

Besides the perfect fluid limit $\mathrm{Kn\rightarrow 0}$, another extreme
behavior is the limit in which the microscopic scales become much longer
than the macroscopic ones, i.e., $\mathrm{Kn\rightarrow \infty }$. In this
case the fluid breaks into free streaming particles. In practice one expects
free streaming to become a good approximation to the evolution when
viscosity becomes so large that $\mathrm{Kn\gtrsim 1}$, see e.g.\ Ref.~\cite{Bouras:2009nn}, 
for a demonstration of a transition from fluid like
behavior to a free streaming behavior. In heavy ion collisions this
transition is called decoupling or freeze-out, which usually is taken to
happen at constant temperature, rather than determined from the condition 
$\mathrm{Kn\sim 1}$, see however, Refs.~\cite{Dumitru:1999ud, Hung:1997du,
Heinz:2007in, Eskola:2007zc, Holopainen:2012id} for a discussion on these
kind of dynamical decoupling conditions.

Besides the Knudsen number, one can quantify the applicability of fluid
dynamics using the inverse Reynolds number, 
\begin{equation}
\mathrm{R}_{\pi }^{-1}=\frac{\sqrt{\pi _{\mu \nu }\pi ^{\mu \nu }}}{P_{0}}%
\text{.}
\end{equation}
In the Navier-Stokes limit, where $\pi _{\mu \nu }\sim 2\eta \sigma ^{\mu \nu }$, 
the inverse Reynolds number becomes proportional to the Knudsen
number. However, in Israel-Stewart theory the shear-stress tensor satisfies
a partial differential equation in which $2\eta \sigma ^{\mu \nu }$ enters
only as a source term. In this sense, the shear-stress tensor does not have
to be equal to the Navier-Stokes value and the inverse Reynolds number
corresponds to another measure to understand the applicability of fluid
dynamics.

In the remainder of this paper, we shall verify how much the fluid-dynamical
description can be driven out of its domain of applicability due to the
large gradients present in a heavy ion collision. Of course, one should
notice that it is basically not known how the breaking of the
fluid-dynamical description will affect the different heavy ion observables.
Such a thing cannot be tested solely within a fluid-dynamical model. In this
sense, the phenomenological implication of our findings will not be fully
addressed in this work.

\section{Fluid dynamical model for heavy-ion collisions}

\label{sec:urhic}

The main ingredients that affect the spacetime evolution and the values of
the Knudsen numbers are the initial energy density profile, initialization
time $\tau_0$ and, obviously, the values of $\eta/s$. In order to better
estimate the uncertainties in calculating the Knudsen numbers, we shall
consider several different choices for these parameters.

Although we consider here a full (3+1)--dimensional expansion, the
longitudinal expansion is treated analytically by assuming longitudinal
boost-invariance. Within this approximation the longitudinal flow velocity
is given by $v_z = z/t$, and the fluid dynamical quantities become
independent of the spacetime rapidity $\eta_{s} = (1/2) \ln\left[(t+z)/(t-z)%
\right]$, i.e., they depend on the transverse coordinates, $\mathbf{x}
=(x,y) $, and the longitudinal proper time, $\tau = \sqrt{t^2-z^2}$, only.
From a numerical point of view, this reduces the (3+1)--dimensional problem
to a (2+1)--dimensional one. In this case it is also enough to give the
initial conditions in the transverse plane only.

For the initial state in AA collisions we use two limits based on the
optical Glauber model, eBC and eWN \cite{Kolb:2001qz}, where the energy
density is proportional either to the density of binary collision (eBC), 
\begin{equation}
\varepsilon_{0}(\mathbf{x}, \tau_0) = C T_A(\mathbf{x}-\mathbf{b}/2)T_A(%
\mathbf{x}+\mathbf{b}/2)
\end{equation}
or to the density of wounded nucleons (eWN), 
\begin{align}
\varepsilon_{0}(\mathbf{x}, \tau_0) &= CT_A(\mathbf{x}-\frac{\mathbf{b}}{2})%
\left[ 1 -\left(1- \frac{\sigma_{NN}T_A(\mathbf{x}+\frac{\mathbf{b}}{2})}{A}%
\right)^A\right]  \notag \\
+& CT_A(\mathbf{x}+\frac{\mathbf{b}}{2})\left[ 1 -\left(1- \frac{%
\sigma_{NN}T_A(\mathbf{x}-\frac{\mathbf{b}}{2})}{A}\right)^A\right],
\end{align}
where $\mathbf{b}$ is the impact parameter, and $T_A(\mathbf{x})$ is the standard nuclear overlap function, 
\begin{equation}
T_A(\mathbf{x}) = \int_{-\infty}^{\infty} dz \rho(\mathbf{x}, z),
\end{equation}
where $\rho(\mathbf{x},z)$ is the nucleon density parametrized as Wood-Saxon
profile and $\sigma_{NN}$ is the total inelastic nucleon-nucleon
cross-section. In order to fully specify the initial conditions, we must
also provide initial values to the fluid velocity $u^\mu$ and the
shear-stress tensor $\pi^{\mu\nu}$. In this work these are always set
initially to zero.

The initial state for pA collisions is taken to be 
\begin{equation}
\varepsilon_{0}(\mathbf{x},\tau_{0})=CT_{p}(\mathbf{x})T_{A}(\mathbf{x}),
\end{equation}
where $T_{p}(\mathbf{x})$ is the nucleon overlap function, parametrized as
Gaussian 
\begin{equation}
T_{p}(\mathbf{x})=\frac{1}{2\pi R^{2}}\exp \left[ -\mathbf{x}^{2}/(2R^{2})\right] ,
\end{equation}
with parameter $R=0.43$ fm, chosen to give the rms charge radius of the
proton. Such initial profile should be good enough to estimate the Knudsen
numbers reached in pA collisions.

The proportionality constants are fixed in such way that the multiplicity in
the $0-5$ \% most central AA collisions at RHIC and the LHC is reproduced.
For the p+A collisions we consider several choices that give different final
multiplicities. The initialization time is taken to be either $\tau_{0}=0.2$
fm or $\tau_{0}=1.0$ fm.

\begin{figure}[tbp]
\includegraphics[width=8.0cm]{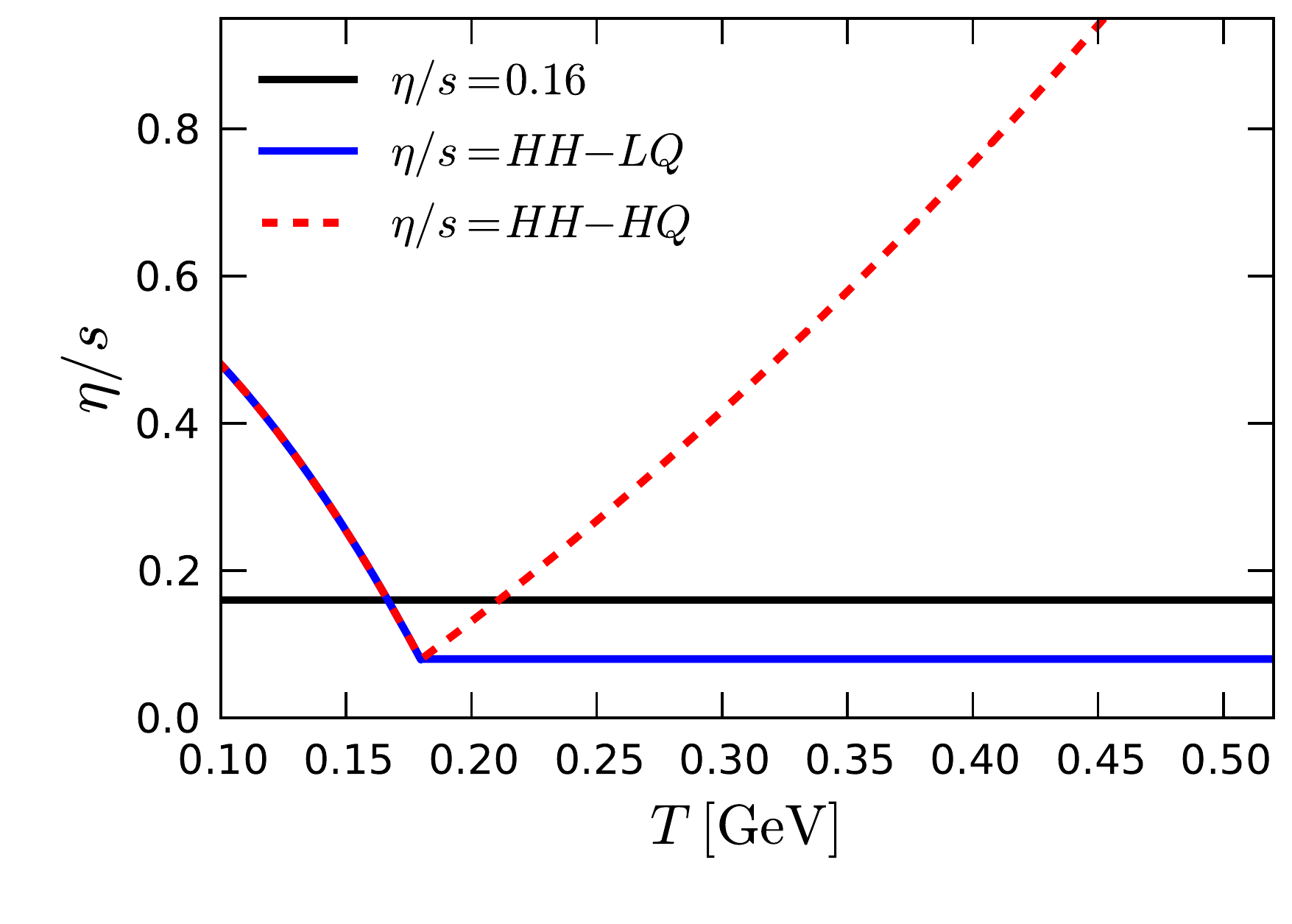} \vspace{-0.5cm}
\caption{(Color online) Different parametrizations of $\protect\eta/s$. }
\label{fig:etapers}
\end{figure}

\begin{figure}[tbp]
\includegraphics[width=8.0cm]{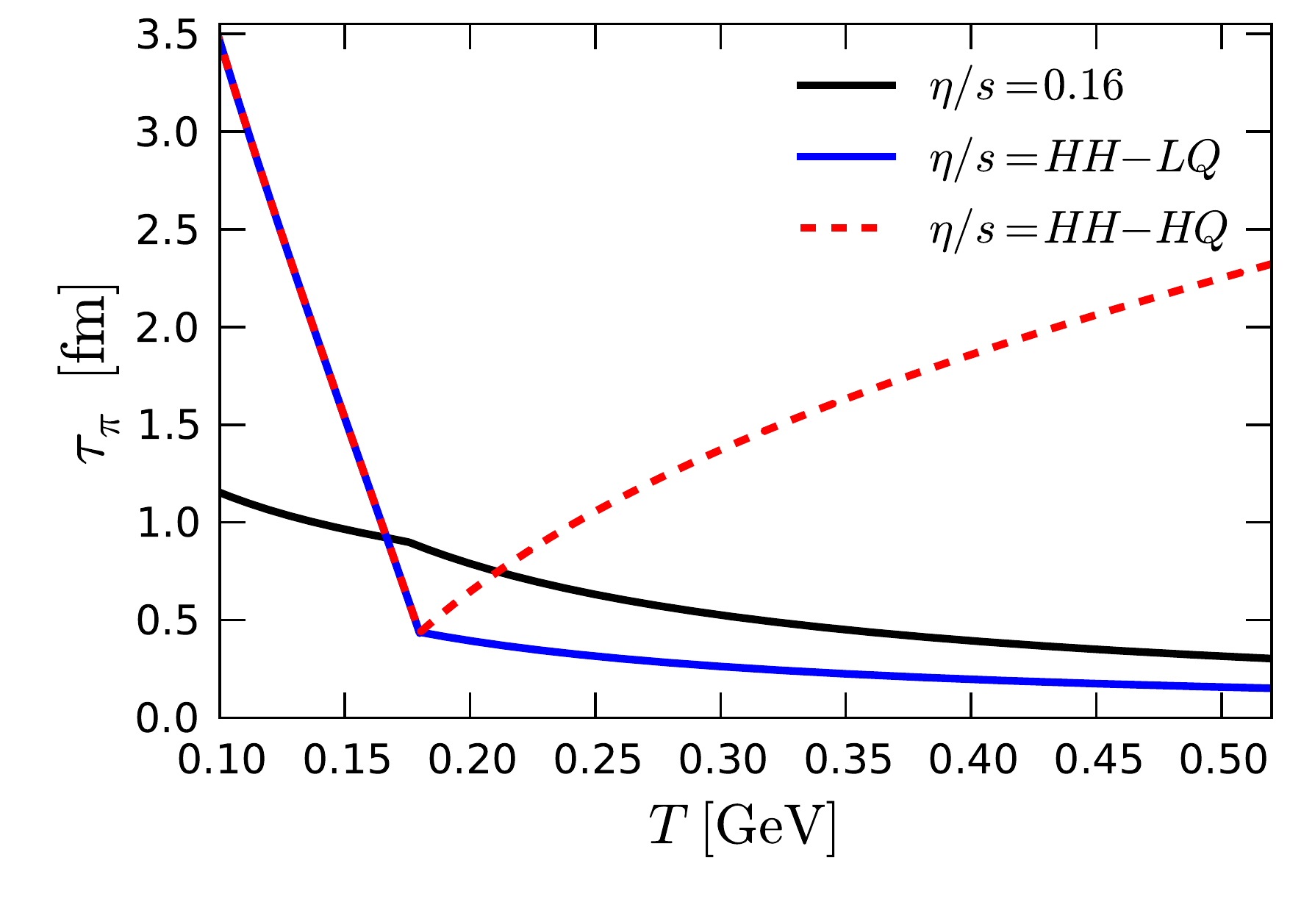} \vspace{-0.5cm}
\caption{(Color online) Relaxation times corresponding to the different $\protect\eta/s$
parametrizations. }
\label{fig:relaxation}
\end{figure}

For the shear viscosity we use three different constant values, $\eta/s =
0.08 $, $0.16$ and $0.24$, and two temperature dependent parametrizations 
\emph{HH-LQ} and \emph{HH-HQ} from Refs.~\cite{Niemi:2011ix, Niemi:2012ry}.
These parametrizations of $\eta/s$ are shown in Fig.~\ref{fig:etapers}, and
the corresponding relaxation times in Fig.~\ref{fig:relaxation}.

Finally, we have to specify an equation of state (EoS) that relates pressure
and temperature to the local energy density. For this we use a lattice QCD
based EoS from Ref.~\cite{Huovinen:2009yb}, with a partial chemical
freeze-out at $T_{\mathrm{chem}}=175$ MeV.

Once the initial conditions, EoS, and the transport coefficients are given,
the equations of motion for shear-stress tensor, Eq.~\eqref{eq:shear_eq},
and the conservation laws, Eq.~\eqref{eq:cons}, form a closed system of
equations, that can be solved numerically. As a result, we obtain the
spacetime evolution of all the quantities appearing in the energy-momentum
tensor and, subsequently, we can calculate any of the macroscopic scales
defined above. The numerical algorithm employed here to solve the equations
of motion is introduced and discussed in Refs.~\cite{Molnar:2009tx,
Niemi:2012ry}.

\section{Results}

\label{sec:results}

\begin{figure}[tbp]
\includegraphics[width=8.0cm]{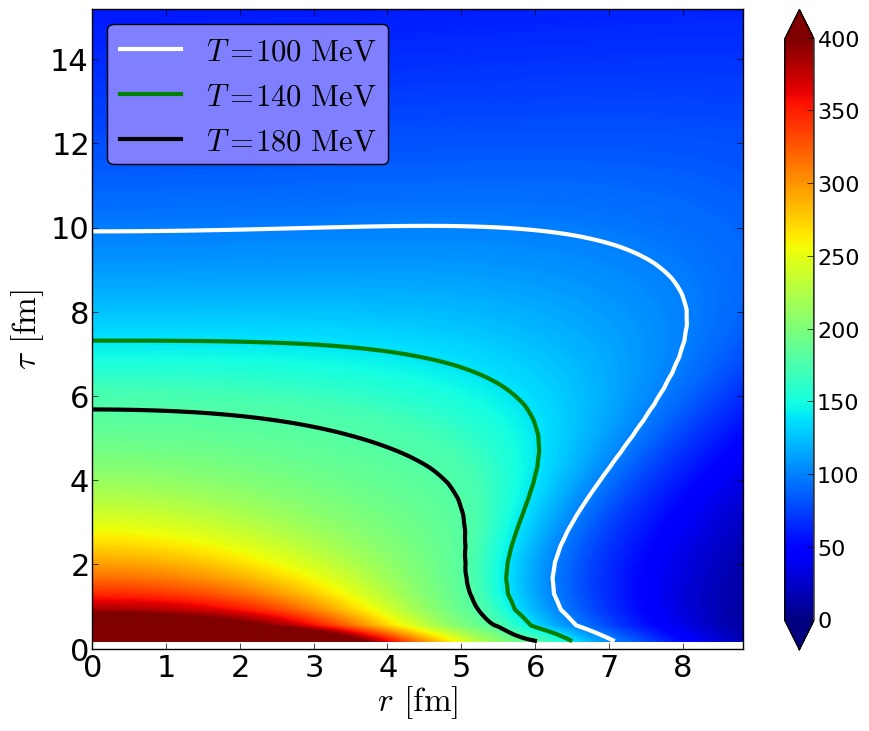}
\caption{(Color online) Spacetime evolution of temperature in $20-30$ \% centrality class
in Pb+Pb collision at the LHC, with $\protect\eta /s=0.08$, $\protect\tau %
_{0}=0.2$ fm and eBC initial state. }
\label{fig:temperature}
\end{figure}

First, we consider the time evolution of semi-peripheral (20-30 \%) Pb+Pb
collisions at the LHC, with different parametrizations of $\eta/s(T)$,
initialization times $\tau_0$ and different initial states. 
Figures~\ref{fig:temperature}-\ref{fig:Reynolds} show the time evolution of different
quantities in the $(r=\sqrt{x^2+y^2},\tau)$--plane along the $x=y$ diagonal.
The impact parameter is always along the $x$-axis. We consider two different
initialization times, $\tau_0 = 0.2$ fm and $1.0$ fm.

In Fig.~\ref{fig:temperature} we show the evolution of temperature with 
$\eta /s=0.08$, $\tau_{0}=0.2$ fm and eBC initialization. Initially the
temperature drops quickly due to the fast longitudinal expansion. In the
later stages of the evolution the longitudinal expansion rate gradually
decreases while, at the same time, the transverse velocity starts to build
up, pushing the matter into the outward direction. This generates a
characteristic shape for the constant temperature curves in the $(r,\tau )$-plane. 
This figure also shows that a significant part of the evolution
happens in the QCD transition region $T=150$--$250$ MeV, which is also the
temperature region where the effects of shear viscosity on the evolution of
the system is the strongest \cite{Niemi:2012ry}. 

\subsection{Spacetime evolution of the Knudsen numbers}

\begin{figure*}[tbp]
\includegraphics[width=8.5cm]{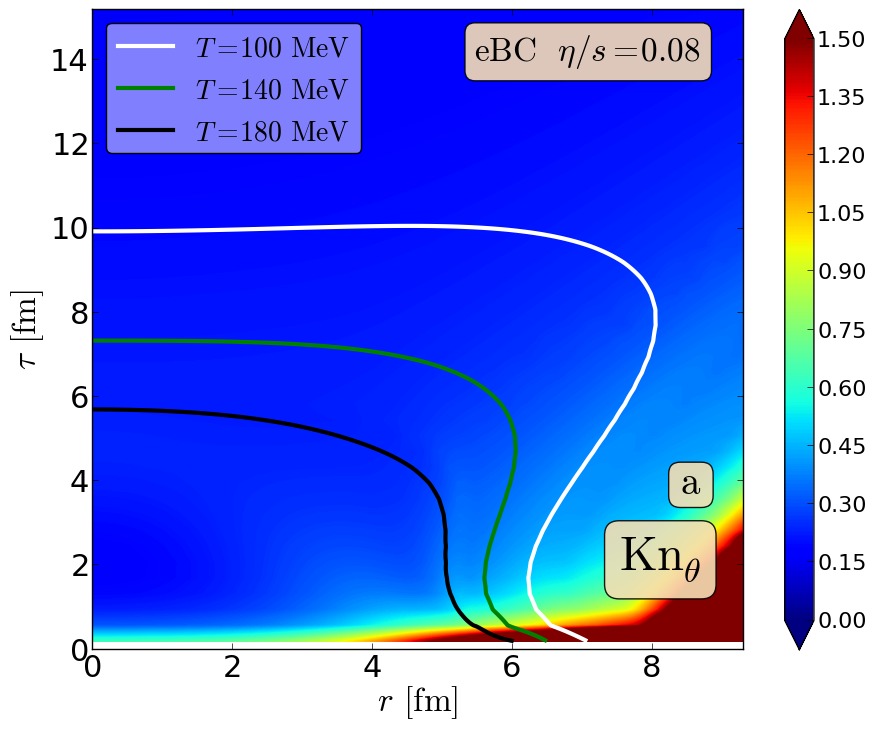}
\includegraphics[width=8.5cm]{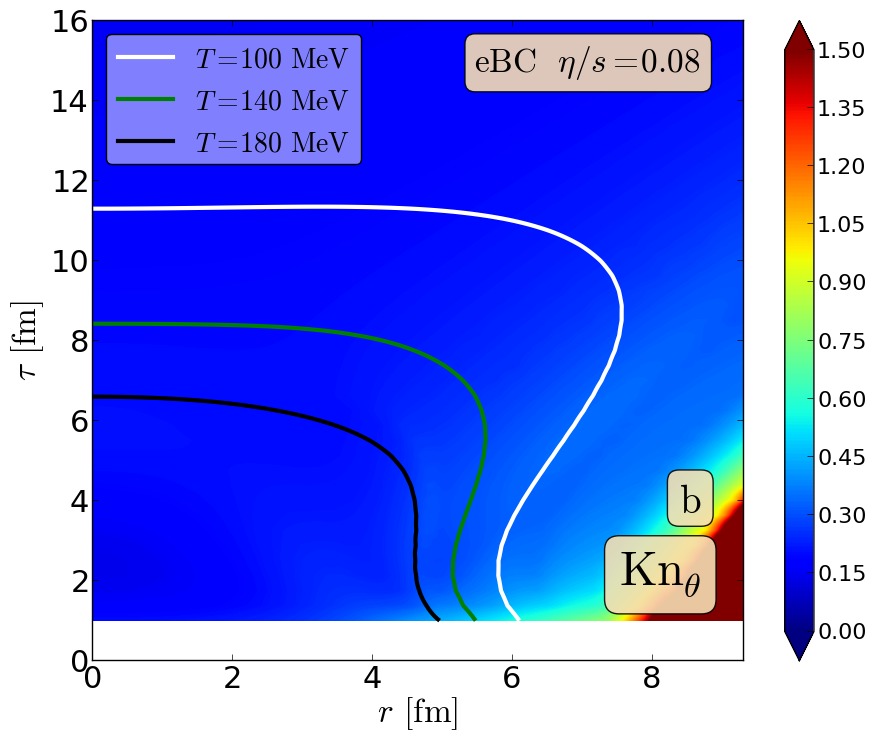}
\includegraphics[width=8.5cm]{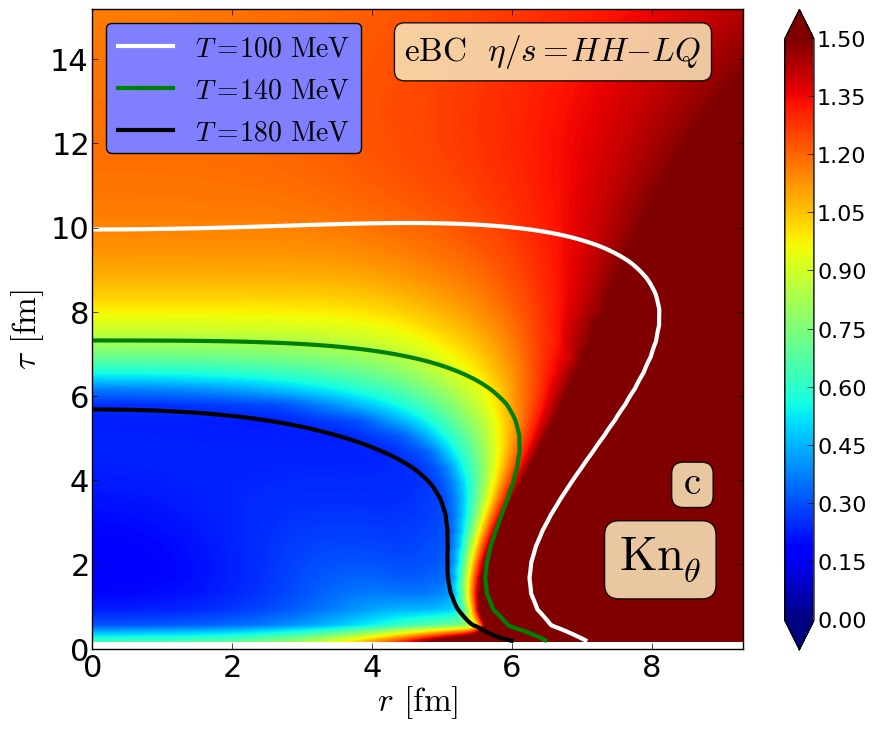}
\includegraphics[width=8.5cm]{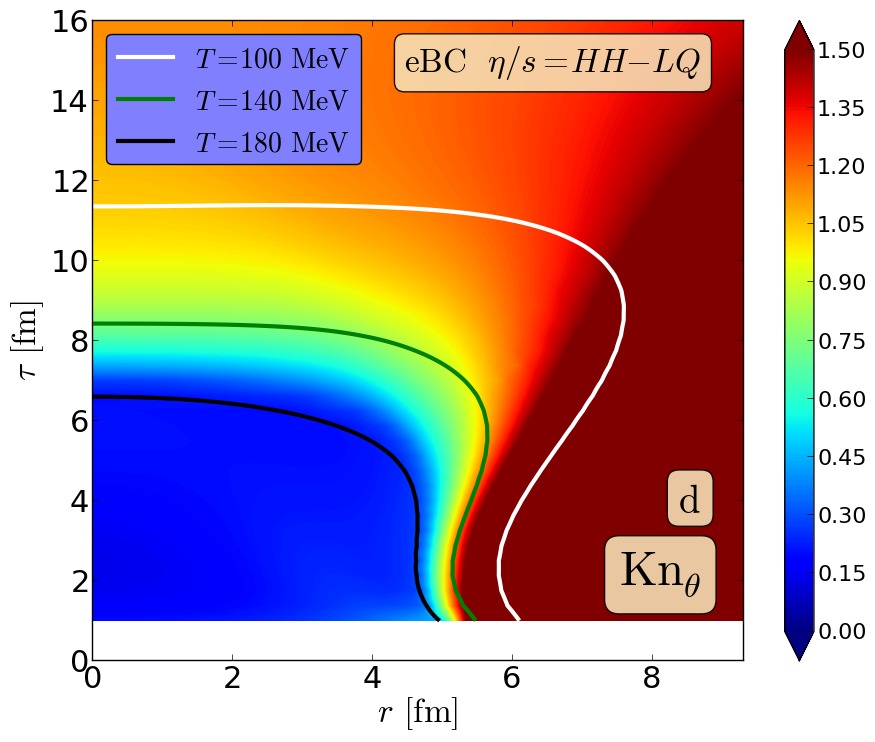}
\includegraphics[width=8.5cm]{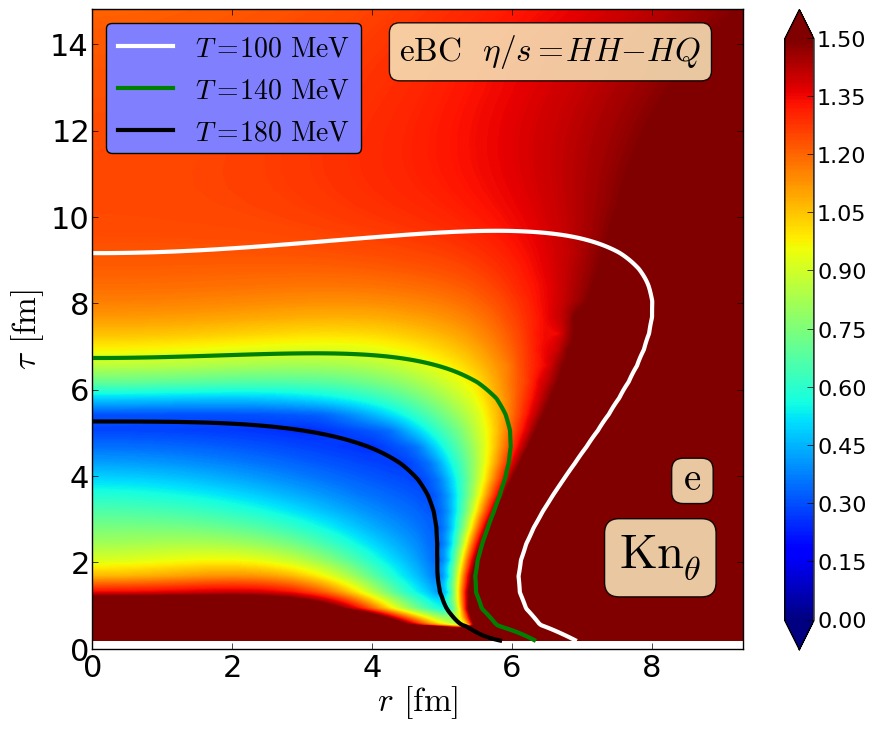}
\includegraphics[width=8.5cm]{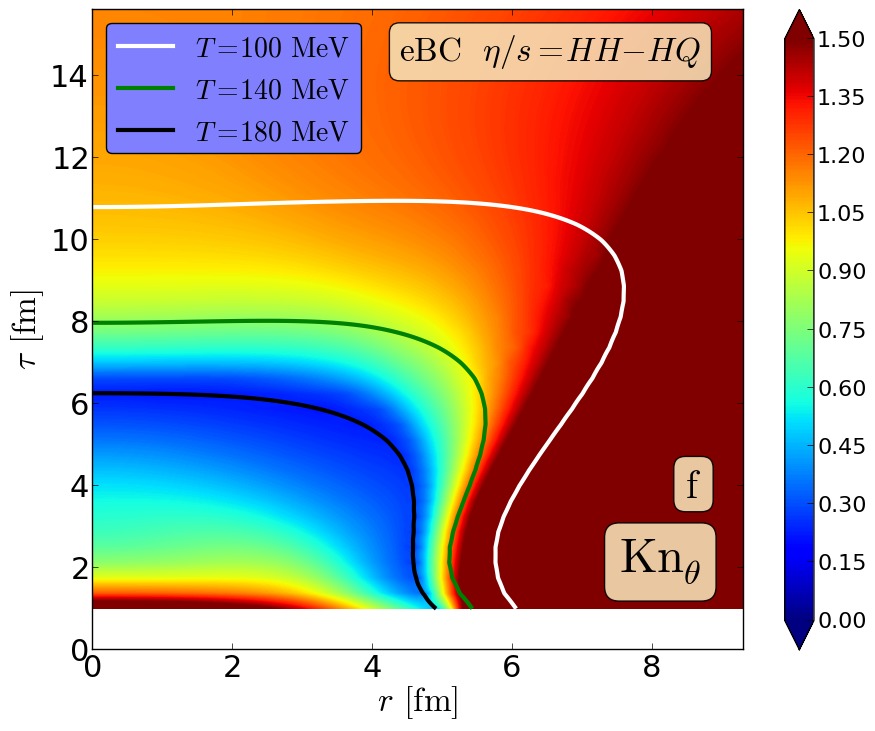}
\caption{(Color online) Spacetime evolution of $\mathrm{Kn_\protect\theta}$ in a Pb+Pb
collision of the $20-30$ \% centrality class at the LHC, with eBC initial
state. In the left panels [(a), (c) and (e)] the initial time is set to $%
\protect\tau_0=0.2$ fm, while in the right panels [(b), (d) and (f)] $%
\protect\tau_0 = 1.0$ fm. In the top [(a) and (b)], middle [(c) and (d)],
and bottom [(e) and (f)] panels, the shear viscosity is set to $\protect\eta%
/s=0.08$, $\protect\eta/s=$\emph{HH-LQ} and $\protect\eta/s=$\emph{HH-LQ},
respectively. }
\label{fig:knudsen_LHC_theta_eBC}
\end{figure*}

\begin{figure*}[tbp]
\includegraphics[width=8.5cm]{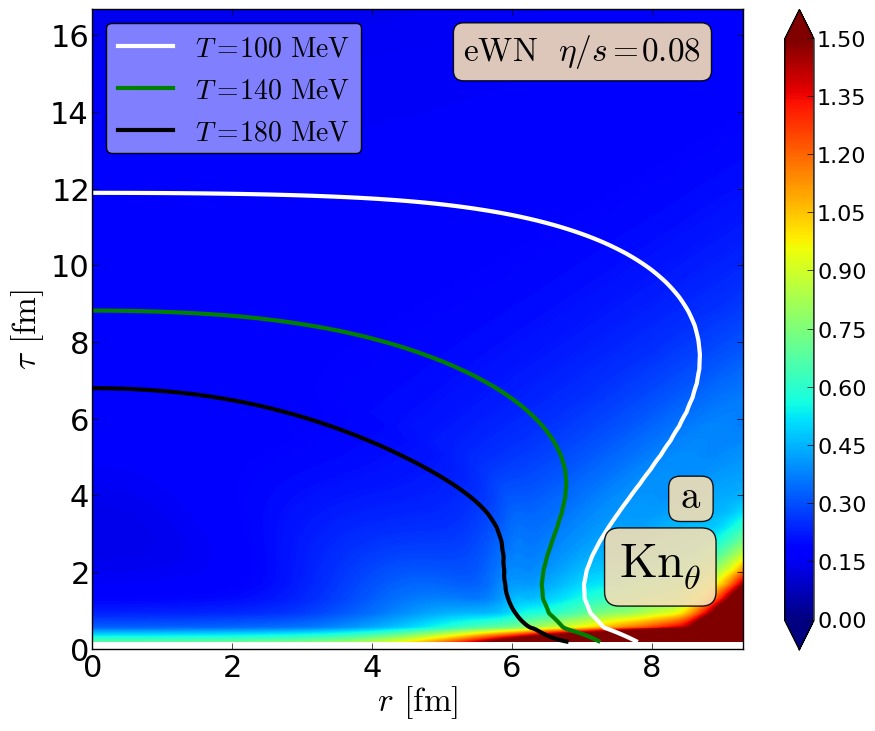}
\includegraphics[width=8.5cm]{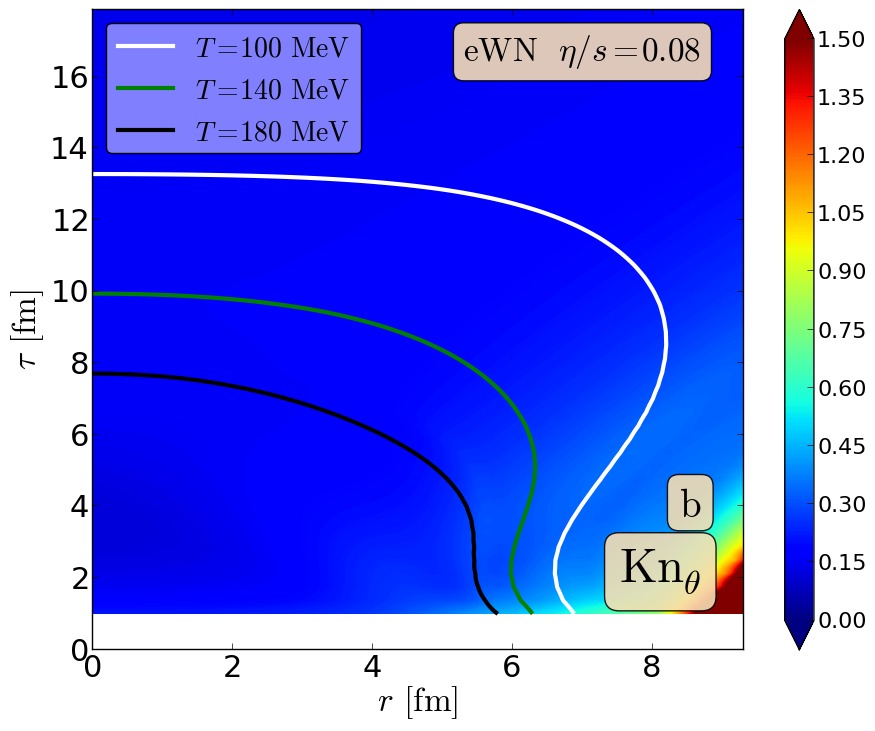}
\includegraphics[width=8.5cm]{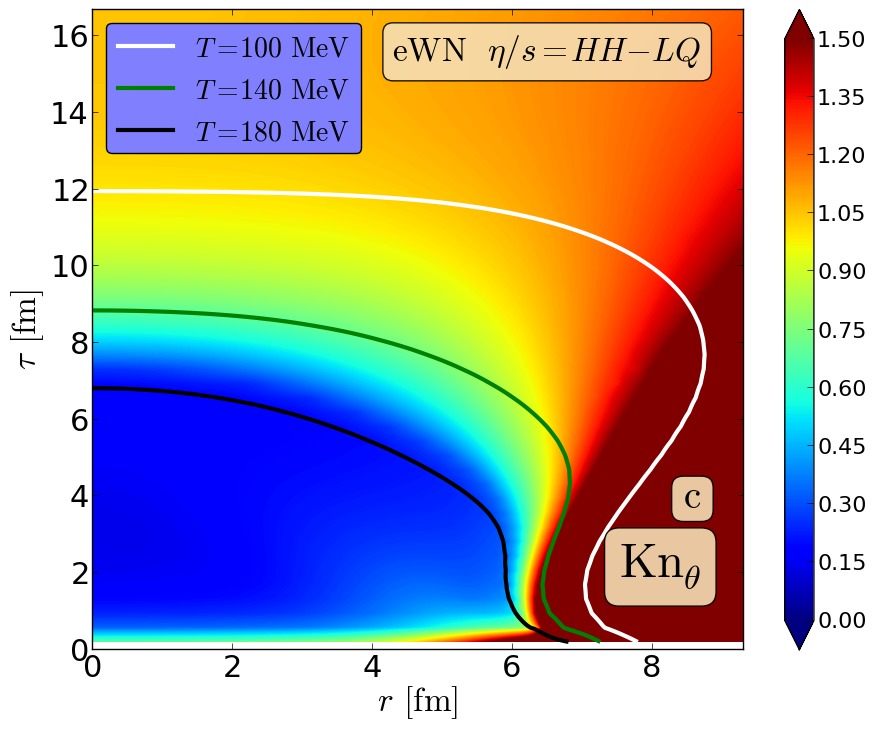}
\includegraphics[width=8.5cm]{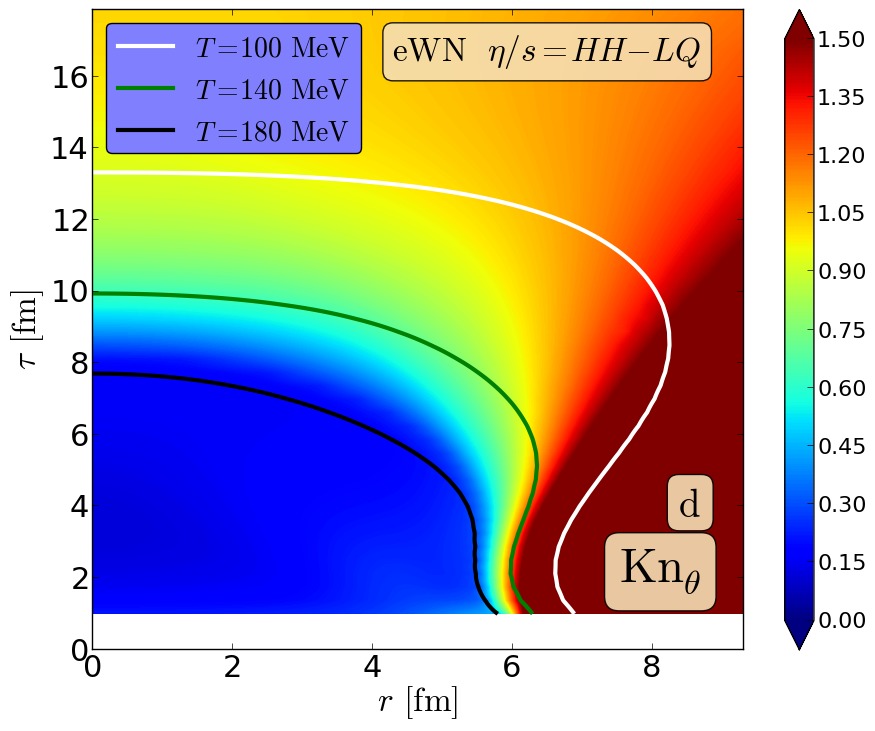}
\includegraphics[width=8.5cm]{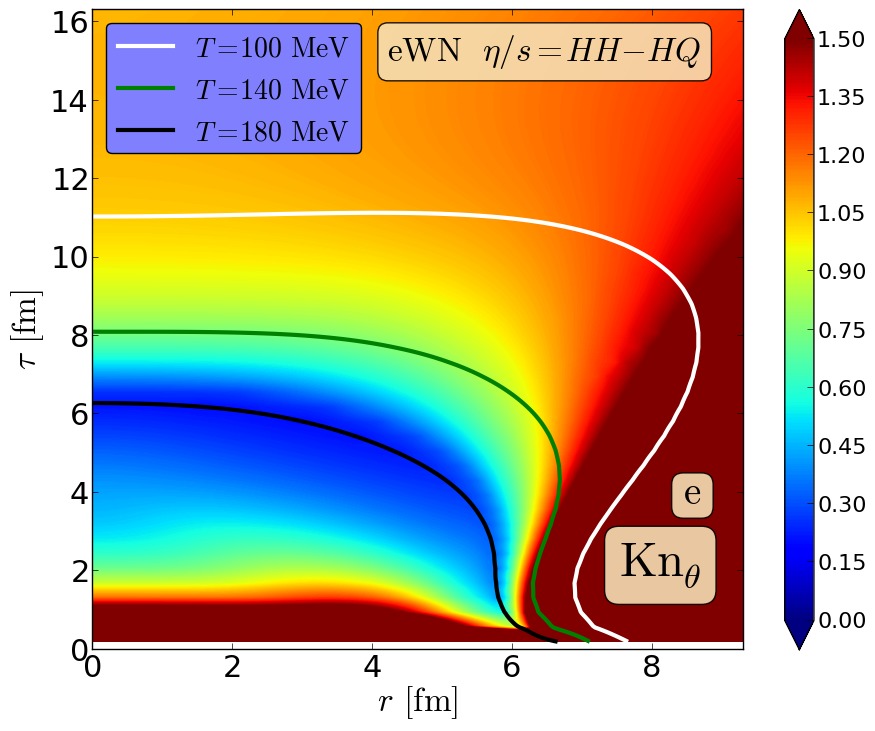}
\includegraphics[width=8.5cm]{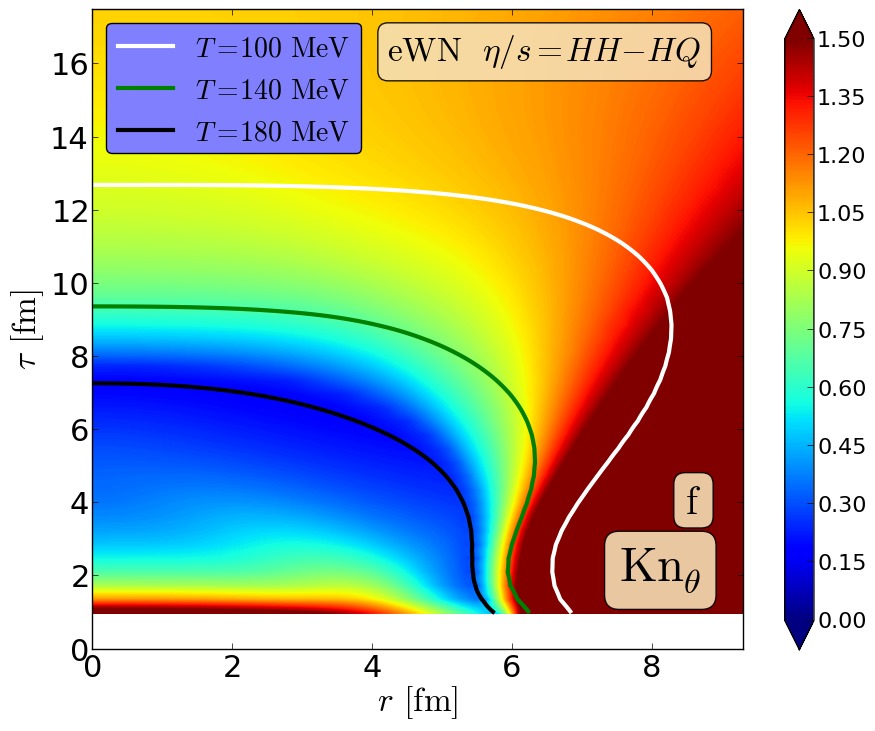}
\caption{(Color online) Spacetime evolution of $\mathrm{Kn_\protect\theta}$ in a Pb+Pb
collision of the $20-30$ \% centrality class at the LHC, with eWN initial
state. In the left panels [(a), (c) and (e)] the initial time is set to $%
\protect\tau_0=0.2$ fm, while in the right panels [(b), (d) and (f)] $%
\protect\tau_0 = 1.0$ fm. In the top [(a) and (b)], middle [(c) and (d)],
and bottom [(e) and (f)] panels, the shear viscosity is set to $\protect\eta%
/s=0.08$, $\protect\eta/s=$\emph{HH-LQ} and $\protect\eta/s=$\emph{HH-LQ},
respectively. }
\label{fig:knudsen_LHC_theta_eWN}
\end{figure*}

\begin{figure*}[tbp]
\includegraphics[width=8.5cm]{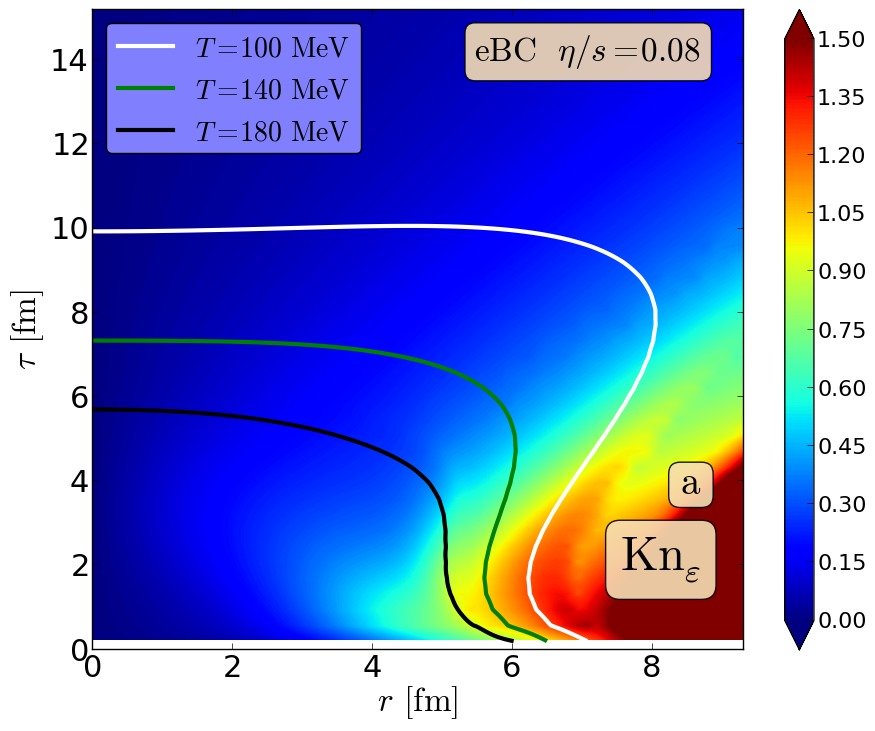}
\includegraphics[width=8.5cm]{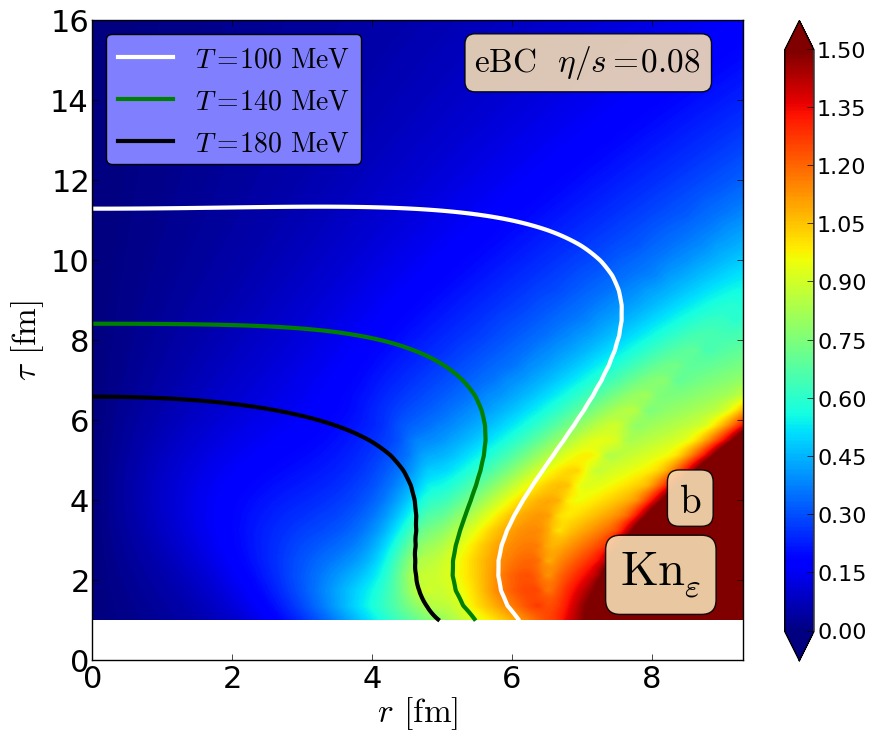}
\includegraphics[width=8.5cm]{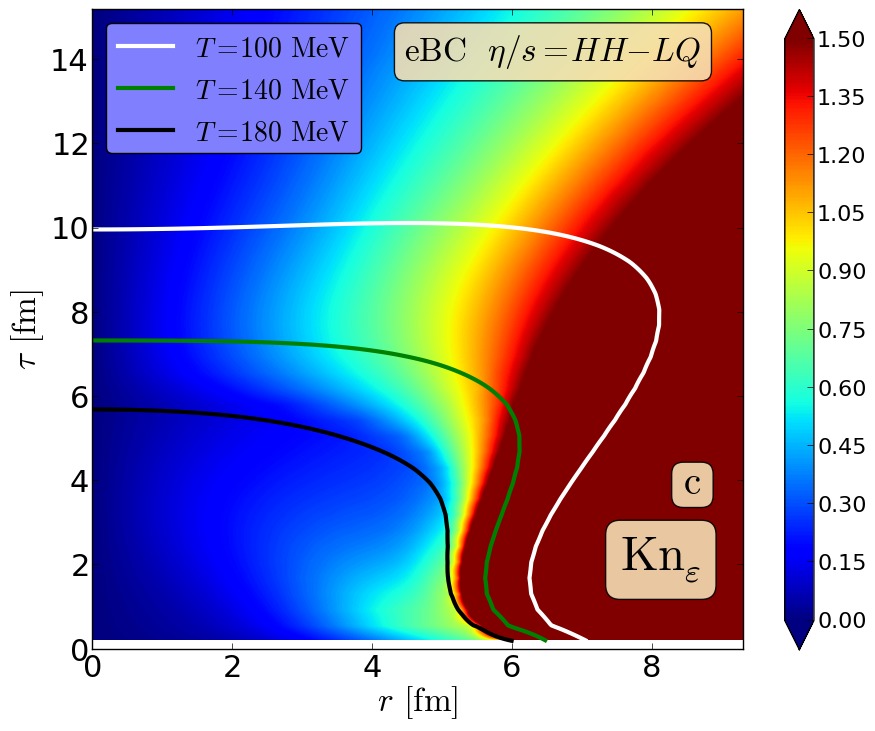}
\includegraphics[width=8.5cm]{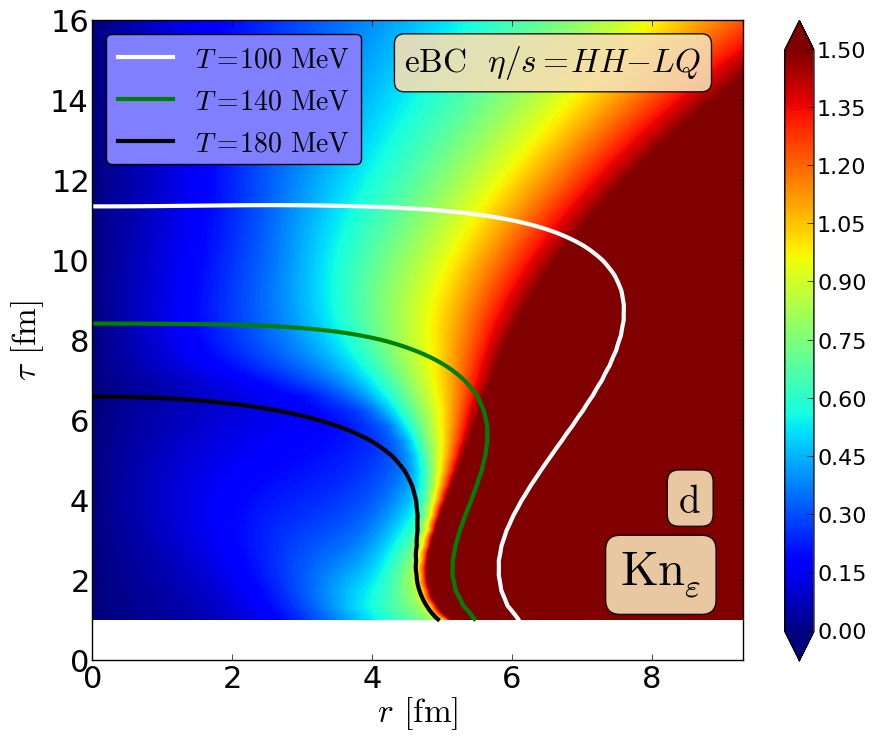}
\includegraphics[width=8.5cm]{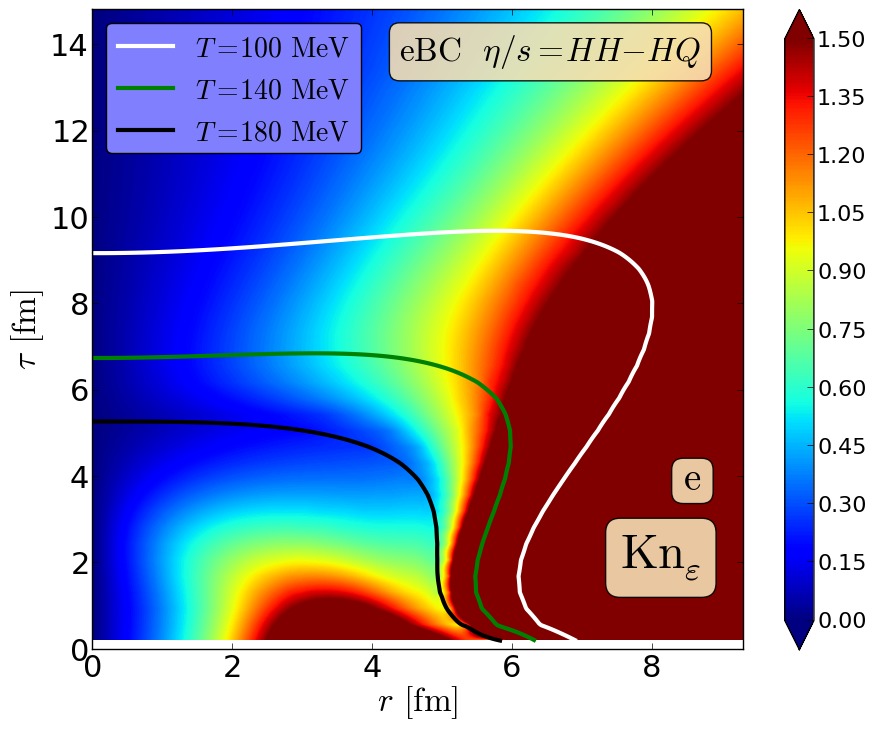}
\includegraphics[width=8.5cm]{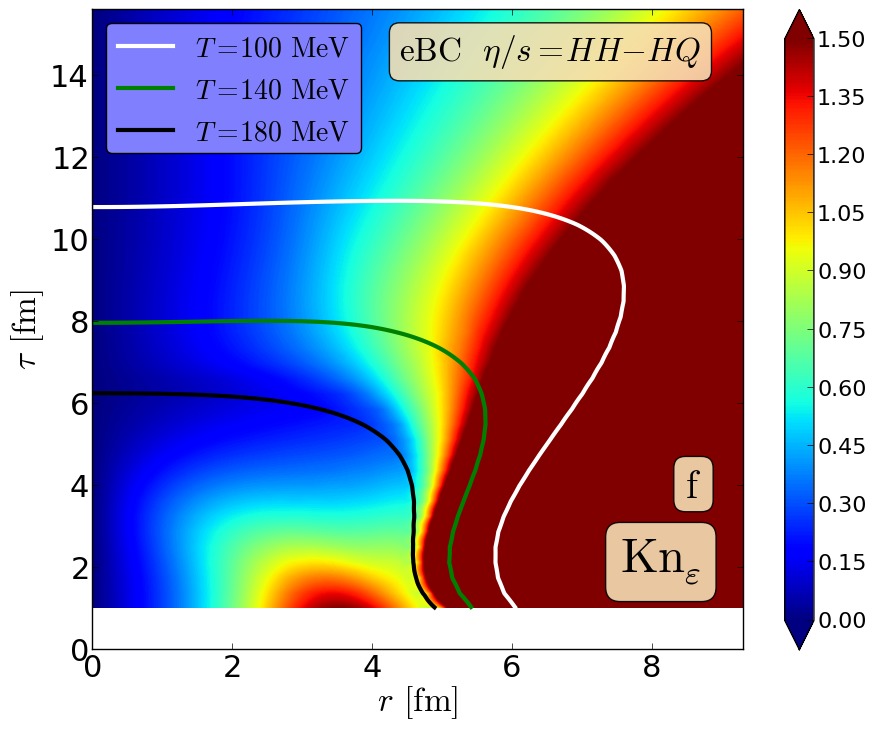}
\caption{(Color online) Spacetime evolution of $\mathrm{Kn_\protect\varepsilon}$ in a Pb+Pb
collision of the $20-30$ \% centrality class at the LHC, with eBC initial
state. In the left panels [(a), (c) and (e)] the initial time is set to $%
\protect\tau_0=0.2$ fm, while in the right panels [(b), (d) and (f)] $%
\protect\tau_0 = 1.0$ fm. In the top [(a) and (b)], middle [(c) and (d)],
and bottom [(e) and (f)] panels, the shear viscosity is set to $\protect\eta%
/s=0.08$, $\protect\eta/s=$\emph{HH-LQ} and $\protect\eta/s=$\emph{HH-LQ},
respectively. }
\label{fig:knudsen_LHC_epsilon_eBC}
\end{figure*}

Figure \ref{fig:knudsen_LHC_theta_eBC} shows the spacetime evolution of $%
\mathrm{Kn_\theta}$ for the eBC initial state, with two different
initialization times, $\tau_0 = 0.2$ fm (Left panels) and $\tau_0 = 1.0$ fm
(Right panels), and three different $\eta/s$ parametrizations, $\eta/s=0.08$
(Top panels), $\eta/s=$\emph{HH-LQ} (Middle panels) and $\eta/s=$\emph{HH-LQ}
(Bottom panels). Figure \ref{fig:knudsen_LHC_theta_eWN} shows the same
cases, but for the eWN initialization. Figure \ref{fig:knudsen_LHC_epsilon_eBC} 
displays the spacetime evolution of $\mathrm{Kn_\varepsilon}$ for the same 
cases as in Fig.~\ref{fig:knudsen_LHC_theta_eBC}. The color coding in the 
figures divides the evolution roughly into three different regions:
\begin{itemize}
\item $\mathrm{Kn<0.5}$ where one expects fluid dynamical behavior (blue).
\item $\mathrm{Kn=0.5\ldots 1}$ a transient region (green to yellow).
\item $\mathrm{Kn>1}$ a free streaming region (red).
\end{itemize}

As already mentioned, it turns out that $\mathrm{Kn_{\theta }}$ and $\mathrm{%
Kn_{\varepsilon }}$ from Eqs.~\eqref{eq:ekspansion} and \eqref{eq:grade}
always give the smallest macroscopic scales and, therefore, are the most
relevant when analyzing the applicability of fluid dynamics in heavy ion
collisions. As can be seen from the figures, the expansion rate is the
dominant scale in the early stages of the evolution, when the longitudinal
expansion is very strong, while the energy density gradient is the dominant
scale at the edge of the fireball. 

While the different choices of initial times and initial conditions have a
quantitative effect on the time evolution of both Knudsen numbers, their
behavior is qualitatively the same for these choices. On the other hand, the
same cannot be said about the shear viscosity coefficient. Even the
qualitative behavior of the Knudsen numbers can change significantly as one
varies this transport coefficient.

For the case of a constant $\eta /s=0.08$, the Knudsen numbers are below the
applicability limit $\mathrm{Kn\ll 0.5}$ almost throughout all the spacetime
evolution, with an exception at the edge of the fireball where the energy
density gradients are very large. At small radius, there is no transition
from a fluid-dynamical regime to a free-streaming regime, except at the very
edge of the system. In other words, for the case of a constant $\eta /s$,
the system never completely switches from a fluid regime to a particle
regime. Therefore, when $\eta /s$ is constant such transition is not
physical, and must be implemented by hand. The common practice in
simulations of heavy ion collisions is just to switch at a constant
temperature hypersurface.

We note that the Knudsen numbers are basically linear with the shear
viscosity coefficient. The larger the shear viscosity, the larger the
Knudsen numbers become. For this reason when using the parametrizations 
\emph{HH-LQ} or \emph{HH-HQ}, which have the larger $\eta/s$ values, except
at the minimum, we observe a drastic increase of the Knudsen numbers.

If we use a realistic hadronic $\eta/s$, which increases as the temperature
decreases, it is possible to observe a transition from a fluid-dynamical
behavior ($\mathrm{Kn < 0.5}$) to a free-streaming behavior ($\mathrm{Kn > 1}
$). Naturally, the $\mathrm{Kn = 1}$ hypersurfaces are never equal to the
constant temperature hypersurfaces. Nevertheless, it is not impossible to
find a case in which temperature stays approximately constant along the $%
\mathrm{Kn = 1}$ hypersurface. For example in Fig.~\ref{fig:knudsen_LHC_theta_eBC}e 
the $T =140$ MeV hypersurface
is very close to the $\mathrm{Kn_\theta = 1}$ hypersurface. So far, we have
not found any example where this happens for a $\mathrm{Kn_\varepsilon = 1}$
hypersurface. 

When using the \emph{HH-LQ} parametrization, which has a constant $%
\eta/s=0.08$ in the high temperature phase, the Knudsen numbers are still
clearly below $0.5$ in the high-temperature phase ($T>180$ MeV), except in
the very early stages of the evolution. If we use instead the \emph{HH-HQ}
parametrization with a strongly increasing $\eta/s$ in the high-temperature
phase, there is only a rather small region around the minimum $\eta/s$ where
fluid dynamics is clearly valid.

\begin{figure*}[tbp]
\includegraphics[width=8.5cm]{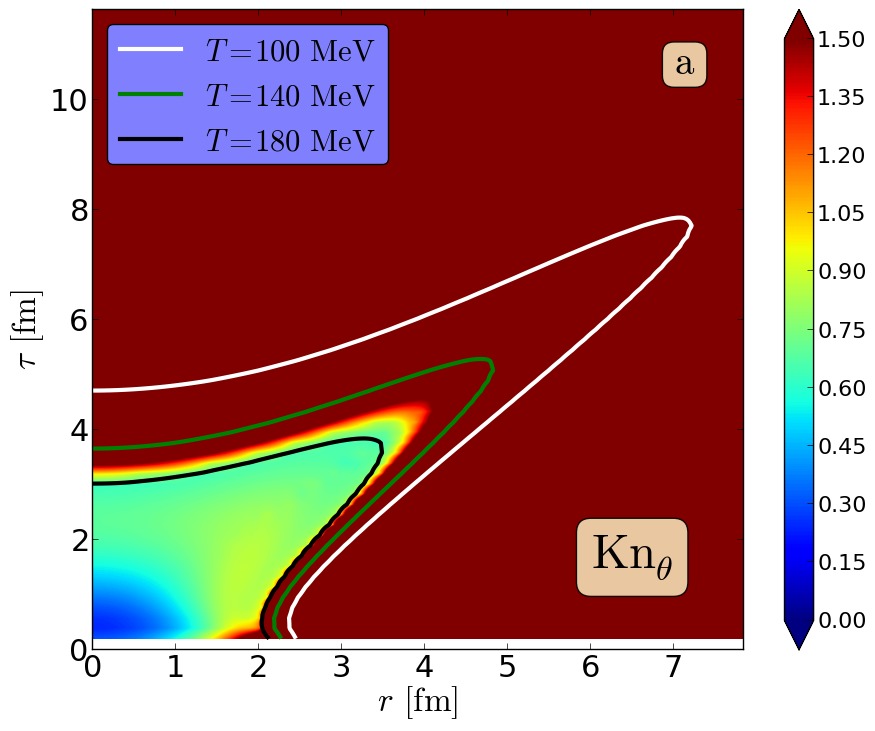} 
\includegraphics[width=8.5cm]{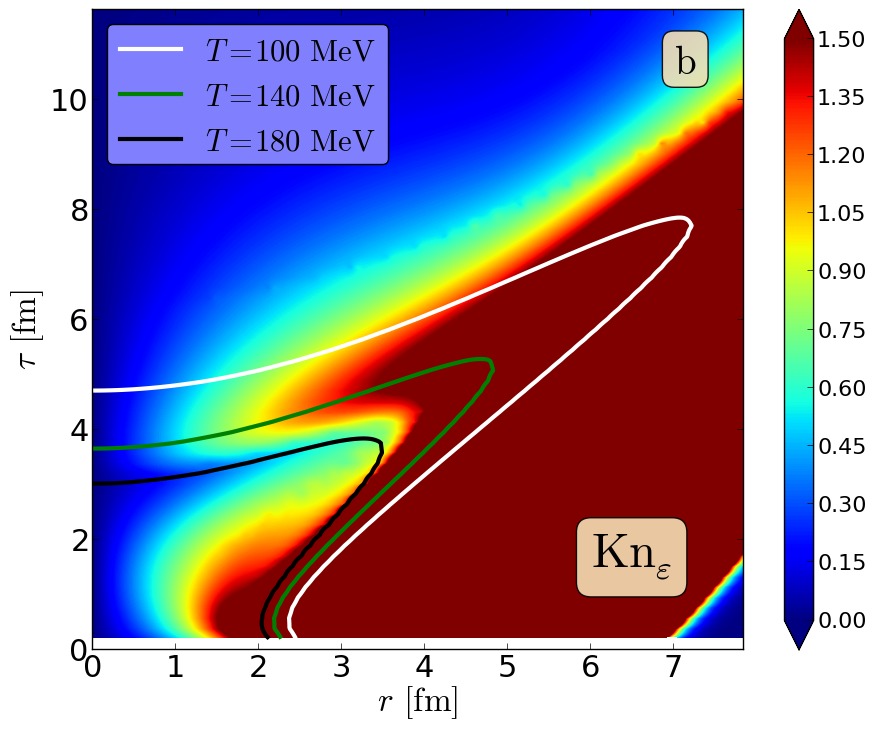}
\caption{(Color online) Spacetime evolution of the Knudsen numbers in p+Pb collision at the
LHC, with $\protect\eta/s=$\emph{HH-LQ} and $dN_{ch}/d\protect\eta = 270$.
(a) $\mathrm{Kn}_\protect\theta$ and (b) $\mathrm{Kn}_\protect\varepsilon$. }
\label{fig:knudsen_pA_etaparam2}
\end{figure*}

In Fig.~\ref{fig:knudsen_pA_etaparam2} we show the time evolution of the
Knudsen numbers in pA collisions using the \emph{HH-LQ} parametrization of 
$\eta/s$. The initial energy density profile is normalized in such a way that
the final charged particle multiplicity is $dN_{ch}/d\eta = 270$. In this
case the lifetime and size of system is considerably smaller than those
achieved in AA collisions. Also, the Knudsen number reached during the
evolution grow considerably faster than in AA collisions: Knudsen number
values at the $T=180$ MeV hypersurface are already large enough to exceed
the $\mathrm{Kn = 0.5}$ limit. As a matter of fact, At the $T=100$ MeV
hypersurface the fluid dynamical description is clearly out of its
applicability domain, with all $\mathrm{Kn_{\theta}}$ values above 1.5. One
can see that in pA collisions the fluid-dynamical description is pushed to
its extreme, even with a constant $\eta/s=0.08$ in the QGP phase. If a
temperature dependent $\eta/s$ were used also in the QGP phase, the
situation would be even more extreme and fluid dynamics would be out of its
domain of applicability even in the early stages of the evolution. We note
that, even though the values of Knudsen numbers are quantitatively different
in pA and AA collisions, the qualitative behavior of the Knudsen number as a
function of spacetime is rather similar in both cases.

\begin{figure}[tbp]
\includegraphics[width=8.0cm]{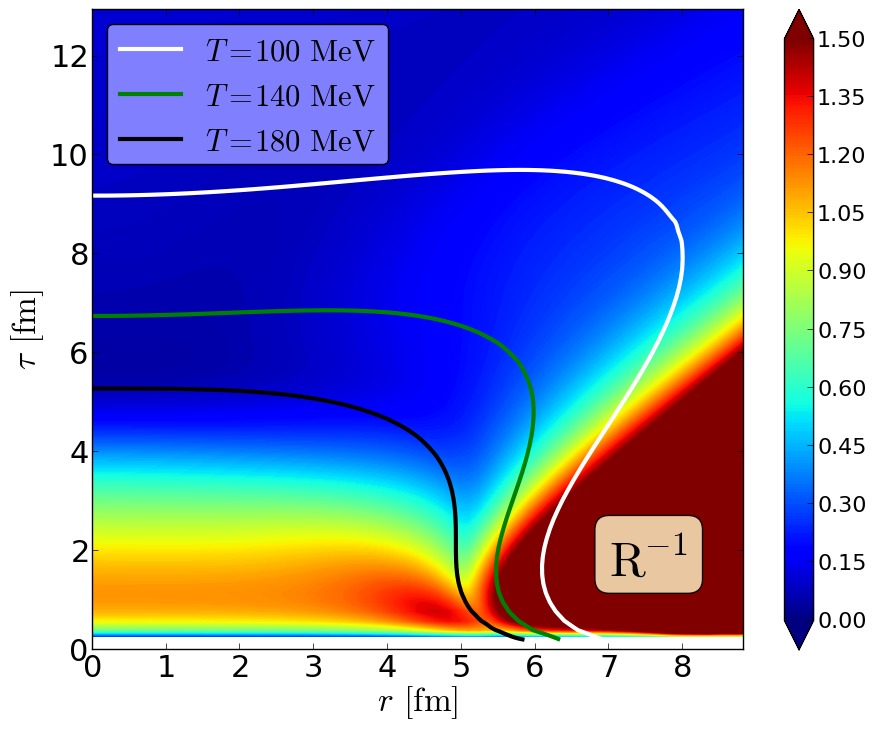}
\caption{(Color online) Spacetime evolution of the Reynolds number in $20-30$ \% centrality
class in Pb+Pb collision at the LHC, with $\protect\eta/s=$\emph{HH-HQ} and
eBC initial state. }
\label{fig:Reynolds}
\end{figure}

So far we have not shown any results involving the inverse Reynolds number.
This is because, the range values of inverse Reynolds number that quantifies
the applicability of the fluid-dynamical description is not very well known.
Commonly, it is accepted is that $\mathrm{R}_{\pi }^{-1}$ should be smaller
than 1, but such limit was never extracted from systematically comparing
solutions of fluid dynamics and Boltzmann equation. We note that the
applicability of the freeze-out formalism currently employed in simulations
of heavy ion collisions depends on the inverse Reynolds numbers, since the
non-equilibrium single particle momentum distribution grows linearly with
the inverse Reynolds number, i.e. 
\begin{equation}
\delta f(k) \sim \frac{\pi^{\mu\nu}}{\varepsilon_0+p_0}k_\mu k_\nu,
\end{equation}
where $k^\mu$ is the 4-momenta of the particle being emitted. In Fig.~\ref{fig:Reynolds} 
we show the time evolution of the inverse Reynolds number for
the same case as in Fig.~\ref{fig:knudsen_LHC_theta_eBC}e. As one can read
off from the plot, the values of inverse Reynolds number become larger than
1 at the edge of the system. From a practical point of view it is difficult
to find a constant temperature hypersurface where the inverse Reynolds
number is always below 1. Note that even by choosing a constant Knudsen
number hypersurface it is not guaranteed that the values of the inverse
Reynolds number would always be less than 1. As mentioned in Sec.~\ref{sec:knudsen}, 
this happens because in transient fluid dynamics the Knudsen
number is not proportional to the inverse Reynolds number, i.e., their
relation is dynamical rather than algebraic.

We note that in Ref.~\cite{Bzdak:2013zma} it was observed that the inverse
Reynolds number is large during a significant fraction of the lifetime 
of the system formed in p+Pb collisions.

\subsection{Maximum effective $\protect\eta/s$}

In this subsection we investigate the systematics of the applicability of
the fluid dynamical description in more detail. We estimate the average
values of $\eta/s$ where the fluid dynamical description breaks down, and
how those values depend on the centrality of the collisions, the initial
conditions and the initialization time. Moreover, we also check what are the
allowed $\eta/s$ values for fluid dynamics to work also in the pA collisions.

For this purpose we first construct an estimate for the maximum $\eta /s$ in
each spacetime point, and then proceed to define a maximum effective $\eta
/s $, for a given initial state and collision centrality, by taking an
average of the local maximum $\eta /s$ over the whole spacetime evolution.

Here we have calculated the Knudsen numbers as 
\begin{equation}
\mathrm{Kn_{i}}=\frac{\tau _{\pi }}{L_{i}}=\frac{5\eta }{\varepsilon_{0}+p_{0}}\frac{1}{L_{i}},
\end{equation}
where $L_{i}$ is one of the macroscopic scales defined above. If the
macroscopic gradients depend sufficiently weakly on the values of $\eta /s$,
as it turns out to be the case, we can calculate $L_{i}$ with any
parametrization of $\eta /s$ in the actual fluid dynamical calculation. This
allows us then to invert the above equation and calculate what is the value
of $\eta /s$ that gives a Knudsen number that is at the limit of the
applicability of fluid dynamics, i.e. $\mathrm{Kn=Kn_{max}}$. In each
spacetime point, this will give a maximum allowed $\eta /s$ for fluid
dynamics to remain valid, 
\begin{equation}
\eta /s|_{\mathrm{max}}=\mathrm{Kn_{max}}\frac{(\varepsilon _{0}+p_{0})L_{i}}{5s}.  
\label{eq:etamax_local}
\end{equation}
Here we take the macroscopic scale to be $L_{i}=\min (L_{\theta
},L_{\varepsilon })$. For each collision system, we then calculate a
maximum effective $\eta /s$ by taking an entropy density weighted average
over the whole spacetime evolution, 
\begin{equation}
\left\langle \frac{\eta }{s}\right\rangle_{\mathrm{max}}=
\frac{\int_{T>T_{f}}d\tau dxdy\tau s(\eta /s|_{\mathrm{max}})}{\int_{T>T_{f}}d\tau
dxdy\tau s},  \label{eq:etamaxaverage}
\end{equation}
where $T_{f}=100$ MeV. The results are not very sensitive to the choice of 
$T_{f}$, e.g.\ $T_{f}=180$ MeV would change the $\eta /s$ limit by order of
10 \% at most. 

In principle, the r.h.s. of Eq.~(\ref{eq:etamax_local})
depends also on the actual values of $\eta /s$ in the fluid dynamical
calculation. However, we have tested that this dependence is weak. We
further note that if a fluid-dynamical calculation is done with a constant 
$\eta /s$, with a value given by Eq.\ \eqref{eq:etamaxaverage}, the resulting
spacetime averaged Knudsen number is approximately $\mathrm{Kn\sim \mathrm{Kn_{max}}}$. 
In this sense the limit $\left\langle \eta /s\right\rangle_{\mathrm{max}}$ 
can be considered as a maximum effective $\eta /s$. As
discussed in Sec.~\ref{sec:knudsen}, we use here $\mathrm{Kn_{max}}=0.5$.

\begin{figure}[tbp]
\includegraphics[width=8.5cm]{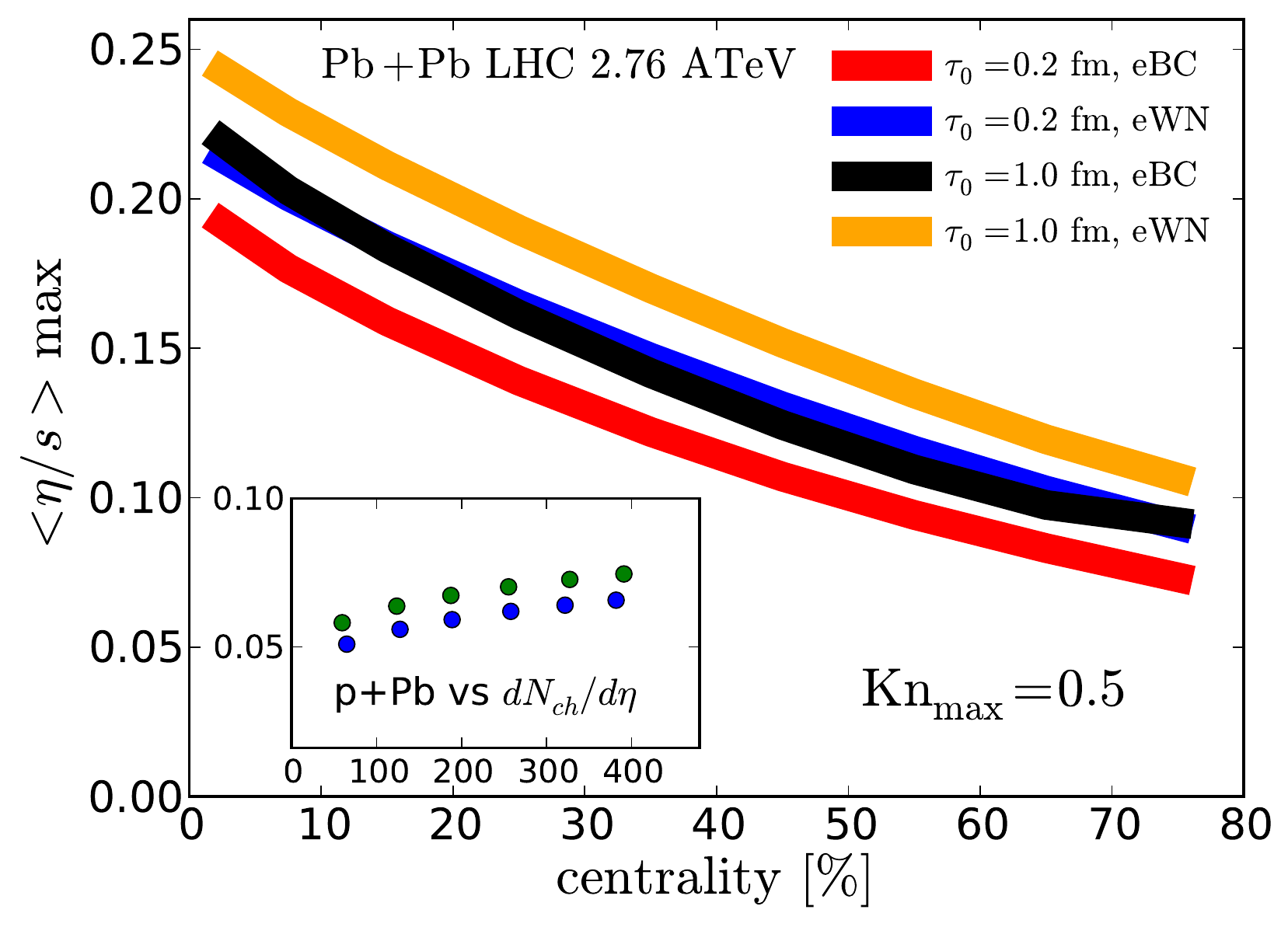} %
\caption{(Color online) The dependence of the $\left<\frac{\protect\eta}{s}\right>_{\mathrm{%
max}}$ on the initial state, initialization time in Pb+Pb collisions at the
LHC. The uncertainty band is obtained by varying $\protect\eta/s$
parametrization, see text. The small figure inside shows the same estimate
for pPb collisions with different multiplicities. The lower (upper) set of
points show the estimate with $\protect\tau_0 = 0.2\,(1.0)$. }
\label{fig:etamaxlimit}
\end{figure}

Figure \ref{fig:etamaxlimit} shows the maximum allowed $\eta/s$ for Pb+Pb
collisions at the LHC as a function of centrality. The different curves
correspond to different choices of the initial conditions and $\tau_0$. The
thickness of the curves indicates the uncertainty of the estimate due to the
actual values of the $\eta/s$ in the fluid dynamical calculation of the
r.h.s. of Eq.~(\ref{eq:etamax_local}). We have varied constant $\eta/s$
between $0.08$ and $0.24$ and also used the two temperature dependent
parametrizations \emph{HH-LQ} and \emph{HH-HQ}. The smaller Figure inside,
where the lower (upper) set of points are for $\tau_0 = 0.2\, (1.0)$ fm,
show the same estimate for pPb collisions as a function of charged hadron
multiplicity.

As can be read off from the figure, the limit for the shear viscosity is of
the order $\eta /s\sim 0.1-0.2$. These values are similar in magnitude to
the estimates for the QGP shear viscosity. As expected, increasing $\tau
_{0} $ or changing initial conditions from eBC to eWN create more favorable
conditions for fluid dynamics to be applicable. Both of these changes have
similar effect on the $\eta /s$ limit.

The collision geometry in AA collisions at RHIC and at LHC is rather similar
and the difference in $\eta/s$ limit between RHIC and LHC is of the same
order as the uncertainty estimate in Fig.~\ref{fig:etamaxlimit}. On the
other hand, a much smaller system formed in pPb collisions results in
drastically smaller values for $\eta/s$ limit, questioning at least
quantitatively how well fluid dynamics can describe such small systems.

\section{Conclusions}

\label{sec:conclusions}

We have calculated the values of different Knudsen numbers in
ultrarelativistic heavy-ion collisions as well as in pPb collisions and,
based on these values, we have estimated the maximum allowed $\eta /s$
values for which the fluid dynamical description of the spacetime evolution
of the collisions is still valid.

We found that two macroscopic scales, the expansion rate and the energy
density gradient, always give the tightest limits for the applicability of
fluid dynamics. The expansion rate is the dominant macroscopic scale during
the early stages of the evolution and near the center of the system, while
the energy density gradients are large near the edge of the system. Other
estimates of the macroscopic scales give typically Knudsen numbers much
smaller than these two choices.

Obviously, the exact values of the Knudsen numbers depend strongly on the
chosen $\eta/s$ parametrization. For example in AA collisions with a
constant $\eta/s=0.08$ the whole evolution is always below the applicability
limit $\mathrm{Kn = 0.5} $, but leads to a rather strange picture where the
system never fully decouples to a free-streaming particles. On the other
hand, the temperature dependent parametrizations considered here give a
decoupling region with a similar shape as constant temperature contours. It
then depends on the high-temperature behavior of $\eta/s$, whether almost
all the QGP evolution is describable with fluid dynamics, or just a small
region around the QCD transition. In pPb collisions the situation is
considerably more difficult: Knudsen numbers are above the applicability
limit almost during the whole evolution even with a small constant 
$\eta/s=0.08$ in the QGP phase.

Based on the calculated values of the Knudsen numbers, we further estimated
the maximum allowed $\eta/s$ for the fluid dynamical description to be valid
for the heavy-ion collisions. The uncertainties in these estimates were
determined by varying the initial conditions, initialization time and the 
$\eta/s$ parametrizations in the actual fluid dynamic calculation. In AA
collisions we found that these estimates give $\eta/s$ limits that are of the same
order than the typical values of $\eta/s = 0.1\ldots 0.2$ found from the
comparisons between the RHIC and LHC data with the predictions of fluid
dynamical models for AA collisions~\cite{Romatschke:2007mq,
Luzum:2008cw,Schenke:2010rr,Gale:2012rq,Song:2010mg,
Song:2011qa,Shen:2010uy,Bozek:2009dw,Bozek:2012qs, Niemi:2011ix,
Niemi:2012ry, Paatelainen:2013eea}. The limits depend on the centrality of
the collision, dropping by approximately a factor of two from central (0-5
\%) to peripheral (70-80 \%) collisions. The difference between RHIC and LHC
is small.

Although, these estimates indicate that fluid dynamics is applicable in
heavy ion collisions, perhaps even at quantitative level, we are still very
close to the applicability limit and there can still be large corrections to
spacetime evolution even in central collisions. Furthermore, we cannot
reliably study simulations with larger values of $\eta /s$ without using a 
description that goes beyond usual fluid dynamics.

The situation for pA collisions is even more difficult, as the $\eta/s$
limit is already clearly below $\eta/s\sim 0.1$ and depends only weakly on
the charged particle multiplicity. As one expects that a same type of matter
is created both in AA and pA collisions, the transport properties of the
matter should also be the same. This makes it hard to trust quantitatively
the predictions of fluid dynamics in pA collisions. We have not considered
event-by-event fluctuations of the initial densities here, but we note that
including those could bring the local conditions in AA collisions closer to
the conditions in pA collisions.

\section{Acknowledgments}

We thank P.~Huovinen and D.~H.~Rischke for comments and discussions.
The work of H.~Niemi was supported by Academy of Finland, Project No.
133005. G.~S.~Denicol is supported by a Banting Fellowship of the Natural
Sciences and Engineering Research Council of Canada.

\end{document}